\documentclass[a4paper,fleqn]{cas-dc}

\usepackage[numbers]{natbib}
\usepackage{amsmath}
\usepackage{graphicx, caption}
\def\tsc#1{\csdef{#1}{\textsc{\lowercase{#1}}\xspace}}
\tsc{WGM}
\tsc{QE}
\tsc{EP}
\tsc{PMS}
\tsc{BEC}
\tsc{DE}

\begin{document}
\let\WriteBookmarks\relax
\def\floatpagepagefraction{1}
\def\textpagefraction{.001}
\shorttitle{MOHGCAA}
\shortauthors{Y. Liu et~al.}

\title [mode = title]{Multi-Order Hyperbolic Graph Convolution and Aggregated Attention for Social Event Detection}            
\author[1, 2] {Yao Liu}[type=author,
                        bioid=1,
                        orcid=0009-0009-3128-7802]
\ead{liuyao@student.usm.my}

\credit{Methodology, Writing – original draft, Writing – review \& editing, Experiment}

\affiliation[1]{organization={Department of Management and Media, The Engineering and Technology College, Chengdu University of Technology},
                city={Leshan},
                postcode={614007}, 
                state={Sichuan},
                country={China}}
\affiliation[2]{organization={School of Computer Sciences, Universiti Sains Malaysia},
                city={Penang},
                postcode={11800}, 
                state={Penang},
                country={Malaysia}}
\affiliation[3]{organization={Department of Art Design,The Engineering and Technology College, Chengdu University of Technology},
                city={Leshan},
                postcode={614007}, 
                state={Sichuan},
                country={China}}               
\author[2]{Tien-Ping Tan}[orcid=0000-0002-4154-4747]
\cormark[1]
\ead{tienping@student.usm.my}
\credit{Conceptualization, Supervision}
\author[3]{Zhilan Liu}
\ead{timothy4247@163.com}
\credit{Data curation, Visualization, Investigation}
\author[1]{Yuxin Li}
\ead{dorissssli2023@163.com}
\credit{Investigation, Data curation}

\cortext[cor1]{Corresponding author}
\begin{abstract}
Social event detection (SED) is a task focused on identifying specific real-world events and has broad applications across various domains. It is integral to many mobile applications with social features, including major platforms like Twitter, Weibo, and Facebook. By enabling the analysis of social events, SED provides valuable insights for businesses to understand consumer preferences and supports public services in handling emergencies and disaster management. Due to the hierarchical structure of event detection data, traditional approaches in Euclidean space often fall short in capturing the complexity of such relationships. While existing methods in both Euclidean and hyperbolic spaces have shown promising results, they tend to overlook multi-order relationships between events. To address these limitations, this paper introduces a novel framework, Multi-Order Hyperbolic Graph Convolution with Aggregated Attention (MOHGCAA), designed to enhance the performance of SED. Experimental results demonstrate significant improvements under both supervised and unsupervised settings. To further validate the effectiveness and robustness of the proposed framework, we conducted extensive evaluations across multiple datasets, confirming its superiority in tackling common challenges in social event detection.
\end{abstract}

\begin{graphicalabstract}
\includegraphics{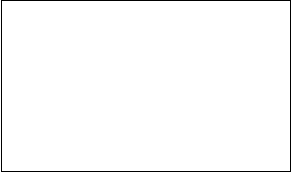}
\end{graphicalabstract}

\begin{highlights}
\item We proposed MOHGCAA, a cutting-edge approach that integrates multi-order hyperbolic graph convolution with aggregated attention. This framework effectively addresses the limitations of traditional Euclidean-based methods by leveraging the hierarchical representation capabilities of hyperbolic space
\item  The proposed framework undergoes rigorous testing across both supervised and unsupervised learning paradigms, demonstrating its adaptability and effectiveness across a range of tasks within hyperbolic space.
\item  Extensive evaluations on a variety of benchmark datasets validate the robustness, scalability, and high performance of MOHGCAA. Its ability to capture multi-order relationships is particularly advantageous for hierarchical and tree-structured data.
\item  MOHGCAA directly encodes higher-order syntactic relationships within hyperbolic space, overcoming the over-smoothing issue commonly associated with deep graph convolution networks. This results in enriched feature diversity and improved performance, especially in scenarios involving complex, tree-like data structures.
\end{highlights}


\maketitle

\section{Introduction}
\label{Introduction}
Social Event Detection (SED) entails the identification of interconnected message clusters from social media data streams to signify certain real-world events. This procedure can be regarded as part of the larger social event data stream. SED is extensively utilized in diverse domains, including sentiment analysis, disaster response, voter trend prediction, and the identification of harmful occurrences. These applications are prevalent in the majority of mobile apps featuring social functionalities, encompassing prominent platforms such as Twitter, Weibo, and Facebook. Consequently, social event detection yields significant insights for both enterprises, aiding in the evaluation of consumer preferences, and public services, facilitating assistance during emergencies and disaster management. Due to its extensive applicability, SED has attracted considerable interest and discourse from both academic and industrial domains. 
\begin{figure}
    \centering
    \fbox{\includegraphics[width=0.45\textwidth]{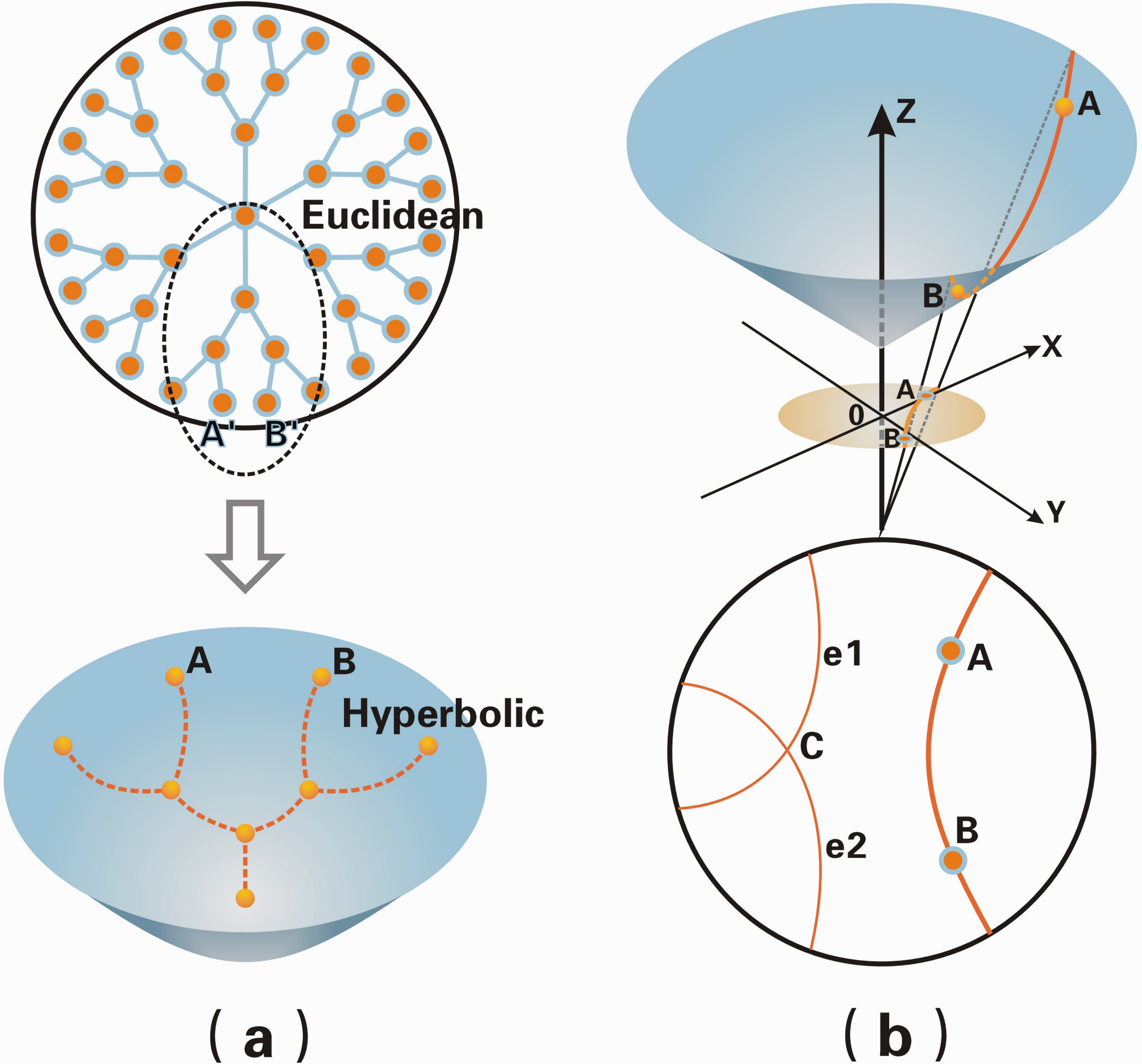}}
    \caption{(a) The points A and B in Euclidean space, situated on distinct branches of the diagram, are in greater
proximity, but points A’ and B’ in hyperbolic space, also on different branches, exhibit a more reasonable closeness to
one another. (b) A diagram illustrating the characteristics of hyperbolic space and its tangent plane, where lines e1 and
e2 are two parallel lines passing through point c, which do not conform to the fifth axiom of contemporary geometry.}
    \label{fig1}
\end{figure}
SED presents unique challenges compared to traditional text classification tasks, primarily due to the nature of social media data. These challenges include short text length, unique data structures, and complex heterogeneous relationships \cite{fedoryszak2019a,peng2019a,zhou2020a,cao2021a,cui2021a,afyouni2022a,qian2023a,qiu2024a}.
\\The first challenge arises from the brevity of social media content. Time constraints, platform settings, or user preferences often constrain posts, resulting in short text that is challenging to process effectively. Creating high-quality labeled datasets from such content is also resource-intensive. Topic-based co-occurrence models were used in the early stages of SED research because they were effective and could show complex word relationships. However, these statistical methods suffered from sparse word representations and overlooked critical semantic information. To fix this problem, more recent research has used unsupervised or self-supervised methods like HISEvnet \cite{cao2024a} and HCRC \cite{guo2024a} to improve performance on high-quality datasets and added semantic context using tools like Word2Vec \cite{mikolov2013a}.
\\The second challenge stems from the tree-like structure of social media data. This structure arises from interactions like replies, mentions, and shares, often influenced by key opinion leaders. In Euclidean spaces, it’s hard to model these kinds of hierarchical relationships well, as shown by how poorly they can capture tree-like patterns (see Figure 1). To overcome this, researchers have turned to hyperbolic spaces, which are better suited for representing hierarchical and tree-structured data. 
\\The third challenge involves the multilayered and heterogeneous relationships inherent in social media. Social interactions generate diverse and complex connections, requiring sophisticated modeling techniques. Meta-path-based approaches, ranging from simple to advanced, have been employed to capture these intricate relationships effectively.
\\Despite the excellent outcomes of current methodologies, they frequently neglect the higher-order interactions among event feature nodes. This research introduces a novel framework named Multi-Order Hyperbolic Graph Convolution with Aggregated Attention (MOHGCAA) to address prevalent issues in social event detection (SED) and the limitations of existing methodologies.
\\The proposed framework leverages the strengths of hyperbolic space to capture hierarchical structures effectively. It begins by projecting node features from Euclidean space into hyperbolic space, capitalizing on its natural suitability for representing hierarchical relationships. In the tangent plane of hyperbolic space at the origin, the framework encodes both first-order and higher-order syntactic relationships simultaneously. This design eliminates the need for excessively deep graph convolutional layers while preserving the richness of learned features. A dynamic attention mechanism further aggregates these multi-order representations, adaptively emphasizing the most task-relevant relationships. The final aggregated features are mapped back into hyperbolic space, ensuring consistency and maintaining the hierarchical integrity of the data for down-stream tasks such as classification.
\\To evaluate the generalizability of our method, we conducted extensive experiments not only on social network datasets but also on multiple general graph neural network datasets. Our framework demonstrated superior performance across diverse tasks. Additionally, we validated its effectiveness in both supervised and unsupervised settings, achieving consistently strong results.
This work makes significant contributions to the fields of Social Event Detection (SED) and hyperbolic representation learning, addressing key challenges and advancing current methodologies.
\begin{enumerate}
    \item Novel Framework: We proposed MOHGCAA, a cutting-edge approach that integrates multi-order hyperbolic graph convolution with aggregated attention. This framework effectively addresses the limitations of traditional Euclidean-based methods by leveraging the hierarchical representation capabilities of hyperbolic space.
    \item Comprehensive Validation: The proposed framework undergoes rigorous testing across both supervised and unsupervised learning paradigms, demonstrating its adaptability and effectiveness across a range of tasks within hyperbolic space.
    \item Strong Experimental Performance: Extensive evaluations on a variety of benchmark datasets validate the robustness, scalability, and high performance of MOHGCAA. Its ability to capture multi-order relationships is particularly advantageous for hierarchical and tree-structured data.
    \item Addressing Higher-Order Dependencies: MOHGCAA directly encodes higher-order syntactic relationships within hyperbolic space, overcoming the over-smoothing issue commonly associated with deep graph convolution networks. This results in enriched feature diversity and improved performance, especially in scenarios involving complex, tree-like data structures.
\end{enumerate}
Through these contributions, this work provides a transformative approach to SED, offering both theoretical and practical advancements in the analysis of hierarchical and heterogeneous social media data.

\section{Relate Work}
\label{RW}

\subsection{Event Detection}
Early event detection research focused on feature engineering to improve model performance. \cite{ahn2006a} utilized constituent and dependency parsers to extract linguistic features like word forms, lemmas, and entity paths, supplemented by contextual information such as semantic roles. POS tagging provided grammatical cues for event triggers, while \cite{liao2011a} demonstrated the utility of unsupervised topic models in uncovering latent connections between events and entities, particularly in addressing imbalanced datasets like ACE-2005. \cite{patwardhan2009a} emphasized contextual features, using lexical and semantic patterns to enhance trigger-entity relationships. Despite their success, these approaches relied heavily on manual feature design, limiting scalability and paving the way for automated neural network methods. The advent of deep learning significantly advanced event detection. \cite{nguyen2015a} introduced convolutional neural networks (CNNs) for event detection, emphasizing feature extraction through pooling techniques. \cite{chen2015a} refined this with dynamic multi-pooling (DMCNN), segmenting sentences around event triggers to enhance semantic representation. Subsequent innovations, such as skip-window CNNs (\cite{zhang2016a}) and parallel multi-pooling CNNs (\cite{li2018a}), improved compositional feature capture. \cite{kodelja2019a} extended CNNs to document-level contexts using a bootstrapping model. However, CNNs often struggled with long-range dependencies, as their localized focus and pooling operations sometimes led to information loss. To address these limitations, recurrent neural networks (RNNs), particularly Long Short-Term Memory (LSTM) networks (\cite{hochreiter1997a}), became prominent. \cite{ghaeini2018a} and \cite{chen2016a} highlighted LSTMs’ ability to model sequential dependencies. \cite{sha2018a} integrated syntactic information using dependency bridges, blending hierarchical structures with sequential modeling. \cite{li2019a} advanced this further with a Tree-LSTM framework enriched by external ontologies, enhancing long-range dependency modeling and event representation. Graph Convolution Neural Networks (GCNs), introduced by \cite{kipf2016a}, extended event detection beyond sequential representations by leveraging graph structures. GCN-ED (\cite{nguyen2018a}) and JMEE (\cite{liu2018a}) used dependency trees to model syntactic relationships. However, reliance on first-order syntactic edges limited these models’ ability to capture diverse relations. \cite{yan2019a} addressed this with the Multi-Order Graph Attention Network (MOGANED), enhancing representation diversity.  \cite{lai2020a} proposed GatedGCN, introducing gating mechanisms to improve layer differentiation, while \cite{cui2020a} developed Edge-Enhanced GCN (EE-GCN) to incorporate dependency label information. Similarly, \cite{dutta2021a} improved event detection by integrating dependency edges in GTN-ED. The progress in social event detection builds upon the advancements in event detection research outlined above. By leveraging these foundational methodologies and adapting them to the unique challenges of social contexts, researchers have refined techniques to identify and analyze events within dynamic, interconnected social environments.

\subsection{Social Event Detection}
Social Event Detection (SED) has emerged as a challenging task that continues to garner substantial attention from researchers. Existing approaches to SED generally fall into two categories: topic-based methods and techniques leveraging heterogeneous information networks (HINs). Topic-Based Methods: Topic-based methods have been widely applied to social media data but face limitations due to the brevity and sparsity of text. For example, Latent Dirichlet Allocation \cite{blei2003a} relies on word co-occurrence statistics, which are often insufficient in short-text scenarios. GPU-DMM \cite{li20161a} mitigates this limitation by integrating Word2Vec with LDA and DMM, adding semantic background information to enhance performance. However, it overlooks relational structures between data nodes. To address this, SGNS \cite{shi2018a} models focus on learning contextual semantic relationships, effectively mitigating keyword sparsity in short texts. Inspired by SGNS, the Semantic-Assisted Non-Negative Matrix Factorization (SeaNMF) model was developed to capture semantics through word-document and word-context relevance in short-text corpora. While these methods have advanced topic modeling for SED, they generally fail to leverage the richness of HINs, which are capable of integrating multiple attribute representations to improve analytical depth. Heterogeneous Information Networks: The application of HINs to SED has shown considerable promise. Hao’s pioneering work introduced an event-based meta-pattern to capture semantic relevance among events, incorporating external knowledge bases to enrich event representation. Building on this, Hao proposed PP-GCN \cite{peng2019a}, a heterogeneous graph convolution network that integrates event metaschemas and external knowledge to achieve fine-grained event categorization. This model’s knowledge-instantiated similarity measure significantly enhanced its ability to identify semantic correlations among events, demonstrating exceptional performance in both detection and clustering tasks. Subsequent research has further exploited graph neural networks (GNNs) for SED. For instance, KPGNN, proposed by \cite{cao2021a}, utilizes incremental learning on HINs to improve detection accuracy. Despite these advances, existing methods struggle to accurately capture the hierarchical and tree-like structures inherent in social media data when constrained to Euclidean space. As highlighted earlier, this limitation poses a significant challenge for modeling multi-level relationships effectively. Our work addresses this gap by adopting hyperbolic space for node representation, enabling the modeling of multi-order relationships while preserving the hierarchical nature of social media data. This novel approach not only resolves structural representation issues but also offers a robust framework for advancing SED methodologies.

\subsection{Hyperbolic Representation Learning}
Recent years have witnessed a surge of research focused on modeling graphs in hyperbolic space. The primary motivation lies in the unique properties of hyperbolic space, which features negative curvature and geometrical advantages well-suited for learning hierarchical representations of graphs. Gromov demonstrated that hyperbolic space is inherently better than Euclidean space for representing tree-like structures. As noted earlier, GCN-based models improve syntactic and semantic modeling compared to conventional neural networks. Researchers \cite{liu2019a} have extended graph neural networks to hyperbolic space by leveraging tangent space. For example, \cite{peng2020a} constructed spatial-temporal graph convolution networks in hyperbolic space for dynamic graph sequences, further exploring projection dimensions within tangent space using neural architecture search (NAS). \cite{elsken2019a} decoupled the message-passing process of GCN into three distinct operations: feature transformation, neighborhood aggregation, and activation. They then adapted each operation to hyperbolic space. Similarly, \cite{bachmann2020a} introduced k-GCN, a mathematically robust generalization of GCN for constant curvature spaces.
In social event detection, hyperbolic space has been applied primarily to social event detection. \cite{qiu2024a} demonstrated success using a simple HNN combined with MLP for social event detection. Building on this, \cite{qiu2024b} proposed the GraphHAM model, which integrates hyperbolic space with an automatic meta-path selection mechanism for heterogeneous information graphs. This framework optimizes meta-path weights and converts them into vectors, significantly reducing the reliance on labeled data. Additionally, the authors designed a Hyperbolic Multi-Layer Perceptron (HMLP) to better capture semantic and structural information in social media contexts.
Despite these advancements, current studies often overlook multi-order representations in hyperbolic space, leaving room for further innovation in event detection.

\section{Preliminaries}
\label{Perliminary}
This study explores the use of hyperbolic space to capture hierarchical relationships in tree-structured data, highlighting its advantages in node classification tasks. To set the stage, this section introduces foundational concepts in graph structures and hyperbolic geometry.
\begin{table*}[]
	\centering
    \caption{The notation used in this work.}
	\begin{tabular}{ll}
		\hline
		Symbol        & Description \\ \hline
		Space Symbol:              \\
		\(\mathbb{E}\)             & Euclidean space.            \\
		\(\mathbb{L}\)              &   Hyperbolic space is represented by the Lorentz model.           \\
		\(\mathbb{K}\)             &      Hyperbolic space is represented by the Kelin model.       \\
		\(\mathbb{B}\)              &     Hyperbolic space is represented by the Poincaré Ball model.        \\ \hline
		Scalaers:     &             \\
	  \(\mathrm{c}\)          &    The curvature of hyperbolic space (\(\mathrm{c}\) < 0).         \\
	  \(1/\sqrt{|c|}\)             &  The radius of an open n-dimensional ball.           \\
		\(\lambda\)              &    The conformal scaling factor.          \\ \hline
		vectors:      &             \\
		\(\mathbf{v},\mathbf{x},\mathbf{y}\)             &   Euclidean vectors.          \\
        \(\mathbf{x}^{\mathcal{B}}\)             &   The node representation in the Poincaré ball model.           \\
		\(\mathbf{x}^{\mathcal{L}}\)              &  The node representation in the Lorentz model.            \\
          \(\mathbf{x}^{E}\)             &  The Euclidean feature vector before hyperbolic mapping.           \\ \hline
		Matrices:         &             \\
		\( A_{\mathrm{along}} \)             &   The adjacency
matrices with directed edges.        \\
		\( A_{\mathrm{rev}} \)             &   The adjacency
matrices with the reverse direction of \( A_{\mathrm{along}} \).          \\
		 \( A_{\mathrm{loop}} \)              &  The adjacency
matrices of an identity matrix for self-loops.          \\ \hline
		Operations:   &             \\
		\(\Vert \cdot \Vert\)    &    The standard norm of Euclidean.    \\
		\({\langle \cdot, \cdot\rangle}_L\)             &    The Minkowski inner product.          \\ 
		\(\mathrm{diag}([...])_n\)             &    The Minkowski metric.          \\
	\(\exp_{\mathbf{o}}^{c}(\cdot)\)              &   The exponential map on the hyperbolic manifold \(\mathbb{H}_c^d\) with curvature \(c\), computed at the origin \(\mathbf{o}\).          \\
		\( \log_{\mathbf{o}}^{c}(\cdot)\)              &   The logarithmic map in the hyperbolic manifold \(\mathbb{H}_c^d\) with curvature is  \(c\) computed at the origin \(\mathbf{o}\).          \\
		\(\oplus_{c}\)              &   The Möbius addition (or hyperbolic addition) with the curvature of the hyperbolic space (often \(c\) < 0)           \\
		\( \oplus \)                &    The element-wise addition.        \\
		\(\kappa_{i,j}\)              &    A learnable score that quantifies the importance of a neighbor \(j\) to a node \(i\) in the neighborhood aggregation.         \\
		\(\otimes_c^{\mathcal{H}}\)        &  The Möbius (hyperbolic) matrix multiplication operation with the curvature \(c\).           \\
	\(P_{o \to \mathbf{x}^{\mathcal{H}}}^c\)              &     The parallel transport function from  the origin \(o\) to the hyperbolic point \(\mathbf{x}^{\mathcal{H}}\) under curvature \(c\).        \\ \hline
	\end{tabular}
	\label{tab:1}
\end{table*}
\subsection{Definition of Graph Structure}
A graph \(\mathcal{G} = (V, E, \mathcal{N}, \mathcal{E})\) consists of node sets \(V\), edge sets \(E\), node type sets \(\mathcal{N}\), and edge type sets \(\mathcal{E}\). A graph is classified as heterogeneous if \(\left|\mathcal{N} \right| > 1\) or \(\left| \mathcal{E} \right| > 1\), and homogeneous otherwise. Traditional Graph Neural Networks (GNNs) often use Euclidean space to model structural relationships, which can fall short in representing hierarchical information inherent to tree-like structures. Hyperbolic space, with its natural capacity for embedding hierarchical data, provides a more robust alternative, although it may still overlook complex inter-node relationships. This research addresses these limitations by adopting hyperbolic models, with the following section detailing relevant hyperbolic geometry concepts as a foundation.

\subsection{Hyperbolic Models}
\begin{figure}
    \centering
    \fbox{\includegraphics[width=0.45\textwidth]{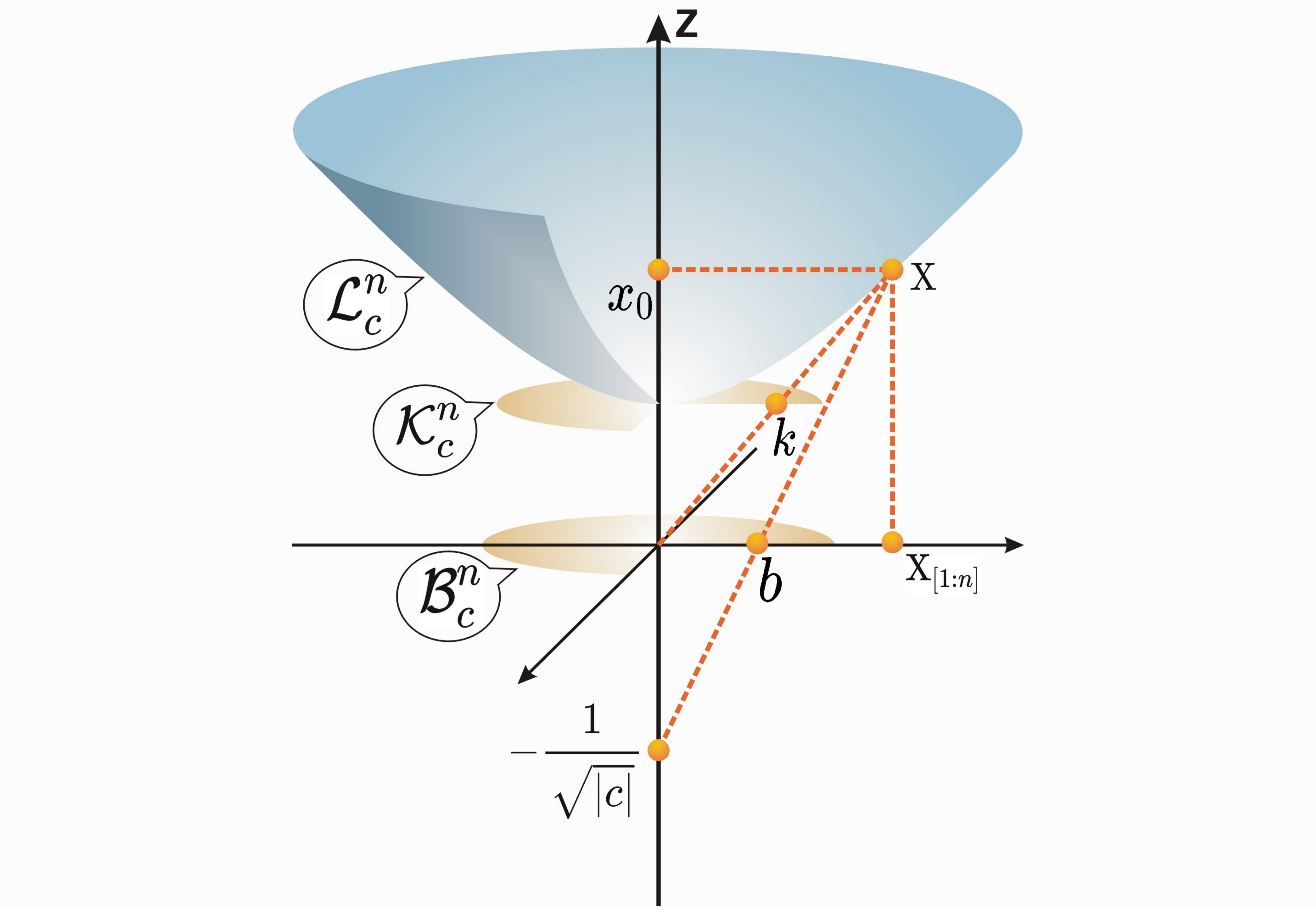}}
    \caption{Depictions of three hyperbolic graphical models: the Lorentz model, the Kälin model, and the Poincaré ball model.}
    \label{fig2}
\end{figure}

Hyperbolic geometry, a type of Riemannian manifold with negative curvature, contrasts with the zero curvature of Euclidean space and the positive curvature of spherical geometry. Several models represent hyperbolic space, including the Poincaré Ball Model \(\mathbb{P}\), the Kälin Model  \(\mathbb{K}\), and the Lorentz Model \(\mathbb{L}\), as illustrated in Figure \ref{fig2}. While each model has distinct mathematical characteristics, they share structural equivalence. The Poincaré Ball Model and Lorentz Model are particularly prevalent in graph representation research, and their definitions are provided below. Here,  \(\|\cdot\|\) denotes the Euclidean norm and \(\langle \cdot, \cdot \rangle_L\) denotes the Minkowski inner product.

Definition 3.1 (Poincaré Ball Model): The Poincaré Ball Model \(\mathbb{P}^n_c\), characterized by negative curvature, is defined as the Riemannian manifold \((\mathcal{B}^n_c, g^\mathcal{B}_\mathbf{x})\), which \(\mathcal{B}^n_c = \{\mathbf{x} \in \mathbb{R}^n : \| \mathbf{x} \|^2 < -1/c \}\) represents an open n-dimensional ball with radius \(1/\sqrt{|c|}\). The metric tensor is given by \(g^B_\mathbf{x} = (\lambda^c_\mathbf{x})^2 g^E\), with \(\lambda^c_\mathbf{x} = 2/(1 + c \| \mathbf{x} \|^2)\) as the conformal factor and \(g^E\) as the Euclidean metric.

Definition 3.2 (Lorentz Model): Also known as the hyperbolic model due to its advantageous visualization properties, the Lorentz Model is defined as the Riemannian manifold \((\mathcal{L}^n_c, g^\mathcal{L}_\mathbf{x})\) with negative curvature, where \(\mathcal{L}^n_c = \{ x \in \mathbb{R}^{n+1} : \langle \mathbf{x}, \mathbf{x} \rangle_\mathcal{L} = 1/c \}\) and \(g^\mathcal{L}_\mathbf{x} = \mathrm{diag}([-1, 1, \ldots, 1])_n\) denotes the Minkowski metric.

\subsection{hyperbolic space representation}
\label{hyperbolicspacerepre}
This section investigates node representations for graph neural networks within hyperbolic space. 
\\In Euclidean space, node features are generally derived from pre-trained embeddings, node-specific attributes, random sampling, or one-hot encodings.  In hyperbolic space, however, these features are projected into either the Poincaré ball model or the Lorentz model to accommodate the curvature of the space.
\\In the Poincaré ball model, the node representation \( \mathbf{x}^{\mathcal{B}} = \exp_{\mathbf{o}}^{c}(\mathbf{x}^{E}) \) involves projecting features onto the tangent space at the origin. This projection is defined by:

\begin{equation}
\label{expP}
\exp_{\mathbf{x}}^{c}(\mathbf{v}) = \mathbf{x} \oplus_{c} \left(\tanh \left(\sqrt{|c|} \frac{\lambda_{\mathbf{x}}^{c} \|\mathbf{v}\|_{2}}{2}\right) \frac{\mathbf{v}}{\sqrt{|c| \|\mathbf{v}\|_{2}}}\right),
\end{equation}

where \( c \) denotes the curvature, \( \mathbf{x} \) is a point in hyperbolic space, and \( \mathbf{v} \) is a Euclidean vector.
\\In the Lorentz model, the mapping is defined as: \( \mathbf{x}^{\mathcal{L}} = \exp^c_o((0, \mathbf{x}^E))\), where a zero is prepended \( \mathbf{x}^E \) to satisfy the condition imposed by the Minkowski inner product. The exponential mapping in the Lorentz model is expressed as:
\begin{equation}
\label{expL}
    \exp^c_\mathbf{x}(\mathbf{v}) = \cosh \left( \sqrt{|c|} \|\mathbf{v}\|_{\mathcal{L}} \right) \mathbf{x} + \frac{\sinh \left( \sqrt{|c|} \|\mathbf{v}\|_{\mathcal{L}} \right)}{\sqrt{|c|} \|\mathbf{v}\|_{\mathcal{L}}} \mathbf{v}.
\end{equation}
\begin{figure}
    \centering
    \fbox{\includegraphics[width=0.45\textwidth]{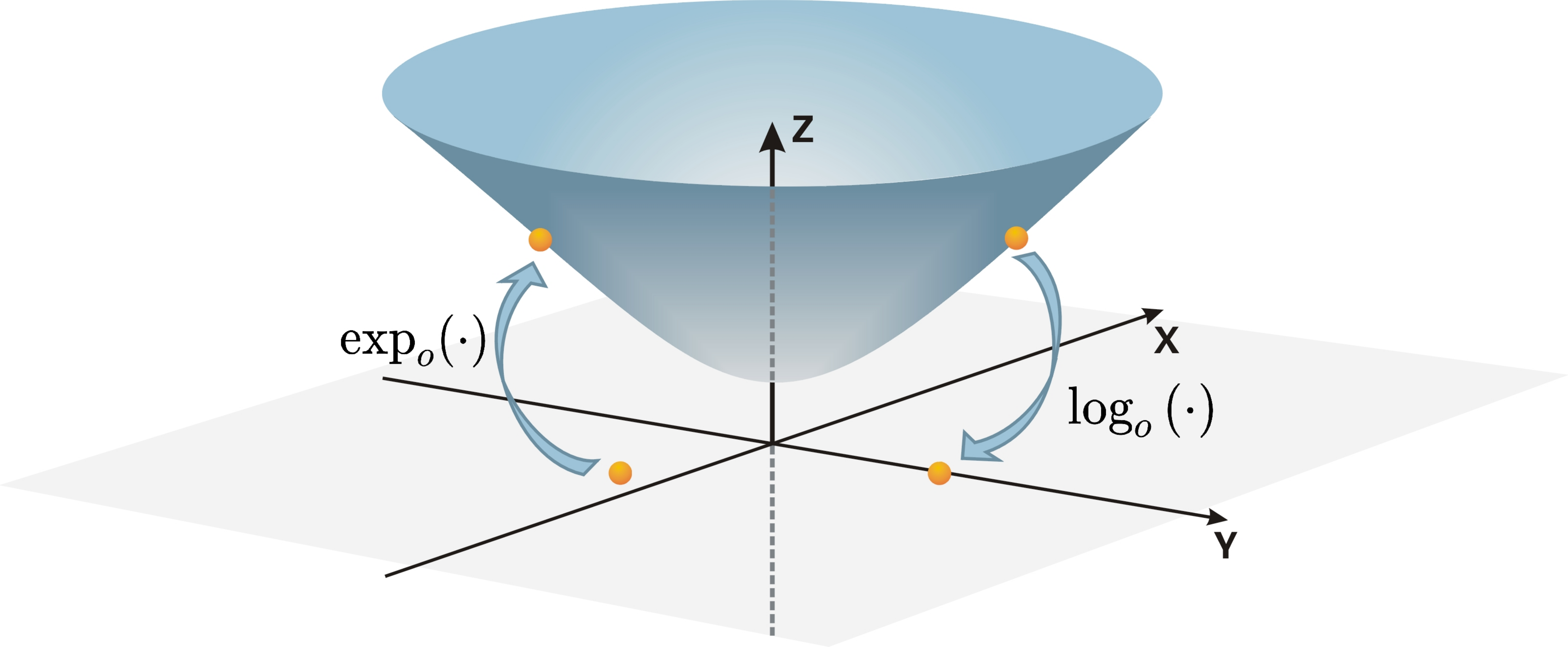}}
    \caption{The \(\exp_o(\cdot)\) operation maps data from Euclidean space to hyperbolic space, whereas the \(\log_o(\cdot)\) operation maps data from hyperbolic space to its Euclidean tangent plane.}
    \label{fig3}
\end{figure}
\\After obtaining the initial hyperbolic space representation, it becomes necessary to perform feature transformation for graph neural networks. The most straightforward method is to complete Euclidean feature operations by projecting the existing hyperbolic space onto the tangent space at a specific point. The notations  \(\log_{x}^{c}(\mathbf{x}^{\mathcal{B}})\) and \(\log_{x}^{c}(\mathbf{x}^{\mathcal{L}}))_{[1:n]}\) representation operations in the Poincaré ball model and the Lorentz model, respectively. 
\\In the case of the Poincaré ball model, the mapping to the tangent space is defined as:
\begin{equation}
\label{logP}
    \log _{\mathbf{x}}^{c}(\mathbf{y})=\frac{2}{\sqrt{|c| \lambda_{\mathbf{x}}^{c}}} \tanh ^{-1}\left(\sqrt{|c|}\left\|-\mathbf{x} \oplus_{c} \mathbf{y}\right\|_{2}\right) \frac{-\mathbf{x} \oplus_{c} \mathbf{y}}{\left\|-\mathbf{x} \oplus_{c}\mathbf{y}\right\|_{2}}
\end{equation}.
\\For the corresponding Lorentz model, the mapping is expressed as:
\begin{equation}
\label{logL}
    \log_\mathbf{x}^{c}(\mathbf{y}) = \frac{\cosh^{-1}(c\langle\mathbf{x}, \mathbf{y}\rangle c)}{\sinh(\cosh^{-1}(c\langle\mathbf{x}, \mathbf{y}\rangle))} (\mathbf{y} - c\langle\mathbf{x}, \mathbf{y}\rangle _\mathcal{L}\mathbf{{x}})
\end{equation}. Figure \ref{fig3} illustrates the mapping procedures between hyperbolic space and the corresponding Euclidean space.After these operations, the resulting features undergo message aggregation and activation. They are then projected back to a new hyperbolic space using the exponential map, yielding a final hyperbolic space-based representation of the graph following a sequence of transformations. For a more comprehensive illustration of the entire process described in this section, please refer to Figure \ref{fig4}.
\begin{figure}
    \centering
    \fbox{\includegraphics[width=0.45\textwidth]{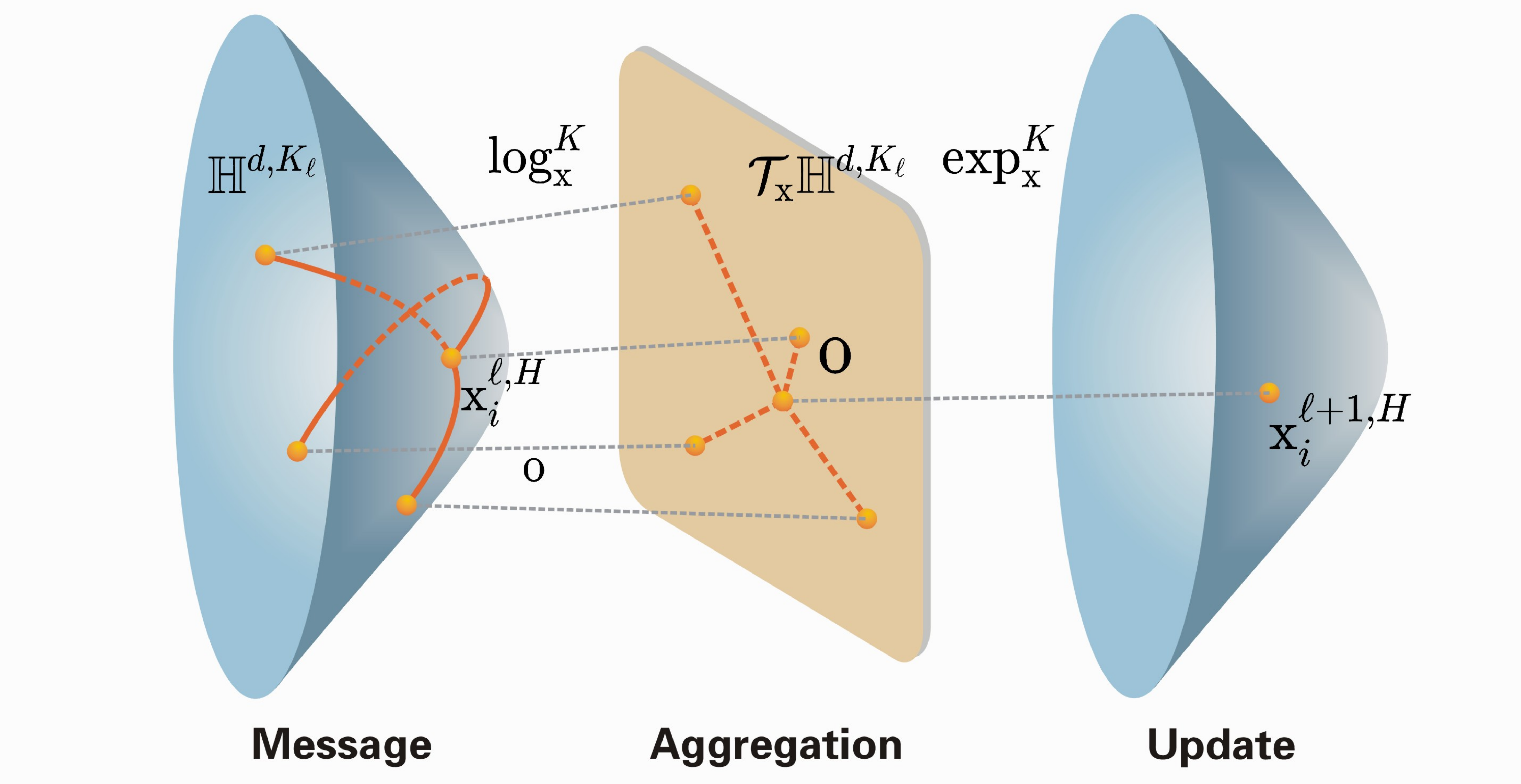}}
    \caption{Upon initialization of the data phase in hyperbolic space, it is projected into the tangent plane of its o-points via the \(\exp_o(\cdot)\) function, and subsequent to the aggregation manipulate, it is implicitly re-mapped into the new hyperbolic space using the \(\log_o(\cdot)\) function.}
    \label{fig4}
\end{figure}

\section{Methodology}
\label{Method}
This section describes our methodology, beginning with an overview of multi-order graph convolution and aggregation in hyperbolic space (Section \ref{MGCA}). We then apply this method to unsupervised learning event detection (Section \ref{UMOHGCA}) and validate its effectiveness in supervised learning settings (Section \ref{SMOHGCA}).

\subsection{Multi-Order Graph Convolution and Aggregated in Hyperbolic Space}
\label{MGCA}
\subsubsection{Overview}
\begin{figure*}
    \centering
    \fbox{\includegraphics[width=0.9\textwidth]{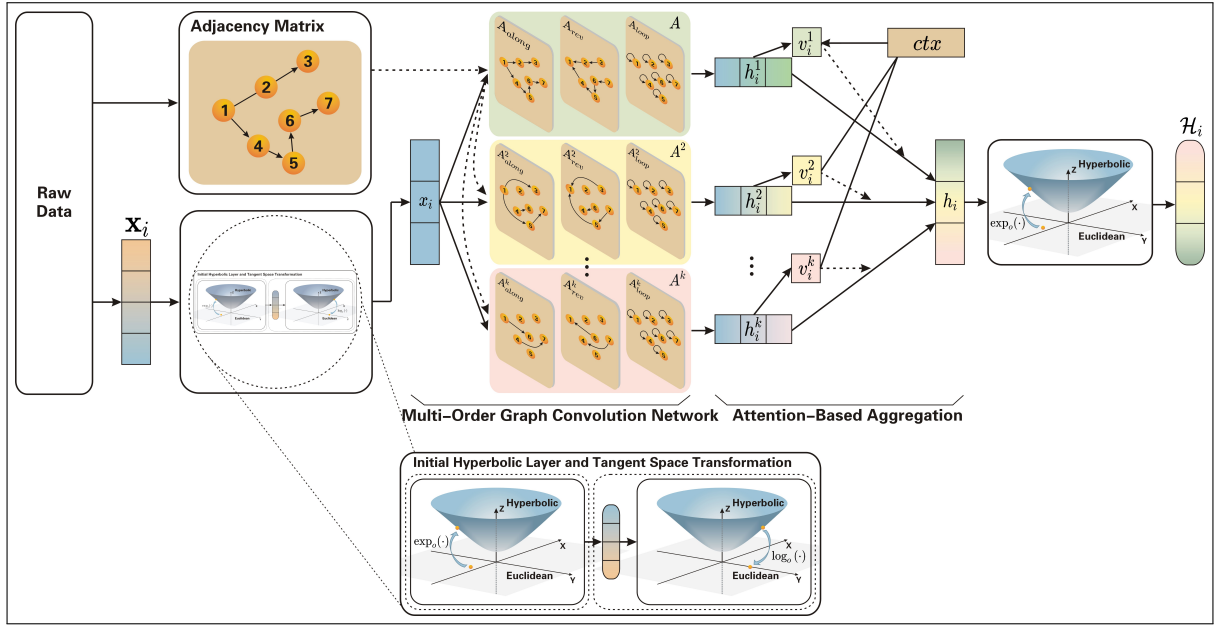}}
    \caption{The overall framework of this study: First, the data representation \(\mathrm{X} _i\) and its corresponding adjacency matrix are obtained. Next, \(\mathrm{X} _i\) is mapped into hyperbolic space via the \(\exp_o(\cdot)\) function, producing its hyperbolic representation. Then, the \(\log_o(\cdot)\) function is applied to project \(\mathrm{X} _i\) onto its tangent space, yielding the representation \(x _i\).On this basis, a multi-order graph convolution network is employed to derive the multi-order representation \(h_i^k\). Subsequently, an attention-based network is used to generate the hyperbolic multi-order graph representation \(h_i\). Finally, the convolution attention representation \(h_i\) is mapped into a new hyperbolic space through the \(\log_o(\cdot)\) function, resulting in the hyperbolic representation \(\mathcal{H} _i\).}
    \label{fig5}
\end{figure*}
To perform multi-order graph convolution aggregation in hyperbolic space, our method follows four key steps: First, we initialize the node features from Euclidean space into hyperbolic space and map these features to the tangent space at the origin. Second, we perform convolution operations of different orders within the tangent space to obtain high-order information about these features. Third, we aggregate the convolution representations from various orders using an attention mechanism. This ensures that the most relevant features are emphasized during the aggregation process. Finally, the aggregated representation in the tangent space is mapped back to a new hyperbolic space using the exponential map. This step obtains in the final output a node feature representation of a high-order convolution aggregation of hyperbolic space.
This structured approach combines the advantages of hyperbolic geometry and graph convolution, enabling efficient and meaningful feature learning. The overall framework can refer the Figure\ref{fig5}.

\subsubsection{Initial Hyperbolic Layer and Tangent Space Transformation}
\label{ihltst}
To initialize node features in hyperbolic space, we project Euclidean node features using the exponential map (Equation \ref{expP} and Equation \ref{expL} in Section \ref{hyperbolicspacerepre}). These features are then mapped to the tangent space at \(o\) via the logarithmic map (Equation \ref{logP} and Equation \ref{logL} in Section \ref{hyperbolicspacerepre}), enabling compatibility with Euclidean operations for graph neural network (GNN) processing.

\subsubsection{Multi-Order Graph Convolution Network}
\label{mogan}
We capture multi-order relationships through adjacency matrices \( A \) for each node, consisting of three submatrices of dimension \( n \times n \): \( A_{\mathrm{along}} \), \( A_{\mathrm{rev}} \), and \( A_{\mathrm{loop}} \).

\begin{itemize}
	\item \( A_{\mathrm{along}}(i, j) = 1 \) if there is a directed edge from \( x_i \) to \( x_j \);otherwise, it is 0.
    \item \( A_{\mathrm{rev}} = A_{\mathrm{along}}^T \) captures the reverse direction.
    \item \( A_{\mathrm{loop}} \) is an identity matrix for self-loops.
\end{itemize}
Our multi-order GCN module aggregates node features across these matrices, yielding feature vectors \( h_i^k \) for each node:

\begin{equation}
\label{multi-orderh}
    h_i^k = f(x_i, a_{along}^k) \oplus f(x_i, a_{rev}^k) \oplus f(x_i, a_{loop}^k)
\end{equation}
where \( f(\cdot) \) is the graph convolution function in hyperbolic space, \(k\) is the orders to aggregate, and \( \oplus \) denotes element-wise addition. The convolution function \( f(p_i, a^k) \) is defined as:
\begin{equation}
\label{convolutionfunction}
    f(x_i, a^k) = \sigma \left( \sum_{j=1}^n \alpha_{ij}^k (W_{a,k} x_j + \epsilon_{a,k}) \right)
\end{equation}
Here, \( \alpha_{ij} \) represents local attention weights based on Ollivier Ricci curvature, computed as:

\begin{equation}
\label{alphaconvolution}
    \alpha_{ij} = \mathrm{softmax}_{j \in N(i)} (\mathrm{MLP}(\kappa_{i,j})),
\end{equation}
where \( \kappa \) captures local structural information. The multi-layer perceptron (MLP) optimizes this computation to incorporate curvature effects.

\subsubsection{Attention-Based Aggregation}
\label{attention}
To aggregate multi-order representations effectively, we apply a weighted attention mechanism:

\begin{equation}
\label{aggregatedatt}
    h_i = \sum_{k=1}^K v_i^k h_i^k,
\end{equation}
where \( v_i^k \) is the attention weight for the \( k \)-order representation of node \( i \), computed as:

\begin{equation}
\label{weightatten}
    v_i^k = \mathrm{softmax}(s_i^k) = \frac{\exp(W_V \tanh(W x_i))}{\sum_{n=1}^N \exp(W_V \tanh(W x_j))}
\end{equation}.
Here, \( W \) and \( W_V \) serve as linear transformations that project node features into scalar attention scores, thereby highlighting the most relevant orders for each node’s representation in hyperbolic space.

\subsubsection{Mapping from Tangent Space to Hyperbolic Space}
After aggregating in the tangent space, the representation is projected back to hyperbolic space using the exponential map (Equation \ref{expP} and Equation \ref{expL} in Section 3.3), aligning the representation with hyperbolic curvature. This projection optimizes the features for downstream tasks such as node classification and link prediction.

\subsection{Multi-Order Unsupervised Hyperbolic Graph Convolution and Aggregated Attention for Social Event Detection}
\label{UMOHGCA}
\subsubsection{Overall Framework}
\begin{figure}
    \centering
    \fbox{\includegraphics[width=0.45\textwidth]{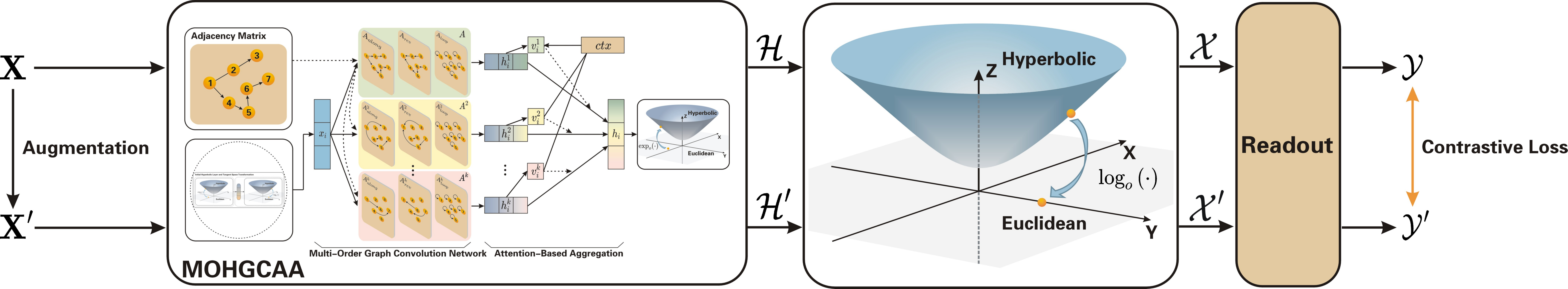}}
    \caption{The framework of the Multi-Order Unsupervised Hyperbolic Graph Convolution and Aggregated Attention for Social Event Detection (MOUHGCAASED) model. \(\mathbf{X} \) represents the obtained node feature, \(\mathbf{X'} \) signifies the augmented node feature of \(\mathbf{X} \), \(\mathcal{H}\) and \(\mathcal{H'}\) indicate the hyperbolic feature following hyperbolic multi-order aggregation, \(\mathcal{X}\) and \(\mathcal{X'}\) denote the Euclidean spatial feature, and \(\mathcal{Y}\) and \(\mathcal{Y'}\) are the ultimate feature representations.}
    \label{fig6}
\end{figure}

High-quality annotated datasets, whether for general or specialized domains, are resource-intensive to create. Contrastive learning, as a self-supervised approach, has gained popularity as it offers a cost-effective alternative. In this work, we employ unsupervised methods to validate the generality, scalability, and robustness of our proposed approach, as depicted in the overall framework diagram (Figure \ref{fig6}). The process consists of three main steps: graph data augmentation, multi-order hyperbolic space node feature graph convolution and aggregation encoding, and contrastive loss optimization.
\subsubsection{Graph Data Augmentation}
Graph data augmentation is essential for contrastive learning, especially in domains like computer vision (CV) and natural language processing (NLP). In the preliminary sections, we introduced the construction of graphs with nodes and relationships. Here, we use graph augmentation techniques to assess how our method affects node representation in hyperbolic space. Specifically, we select feature corruption as the primary augmentation technique for node features. Given a graph \( \mathcal{G} =(V, \mathcal{A}) \) and its augmented version \( \mathcal{G'} =(V', \mathcal{A'}) \), the node set changes (\( V \neq V' \)) while the adjacency matrix remains the same (\( \mathcal{A} = \mathcal{A'} \)).
\subsubsection{Multi-order Hyperbolic Space Node Feature Graph Convolution and Aggregation Encoding}
Following the method we propose in  \ref{ihltst}, we map node features from Euclidean space to hyperbolic space. The transformation is defined as:

\begin{equation}
    h_i^{l, \mathcal{H}} = \left(W^l \otimes_c^{\mathcal{H}} x_i^{l-1, \mathcal{H}}\right) \oplus_{c}^{\mathcal{H}} b^l,
\end{equation}

where \(W^l \otimes_c^{\mathcal{H}} x_i^{l-1, \mathcal{H}}\) represents the hyperbolic tangent space operation described in Equation \ref{expP}:
\begin{equation}
    W \otimes x^\mathcal{B} := \exp_o^c (W \log_o^c (x^B)),
\end{equation}
or alternatively in Equation \ref{expL},

\begin{equation}
    W \otimes x^\mathcal{L} := \exp_o^c (0, W \log_o^c (x^L)[1:n]).
\end{equation}
The bias is  Equation \ref{bais}:
\begin{equation}
\label{bais}
    \mathbf{x}^{\mathcal{H}} \oplus_c^{\mathcal{H}} \mathbf{b}^{\mathcal{H}} = \exp_{\mathbf{x}^{\mathcal{H}}}^c \left( P_{o \to \mathbf{x}^{\mathcal{H}}}^c \left( \log_o^c(\mathbf{b}^{\mathcal{H}}) \right) \right)
\end{equation}
, where \(\mathbf{b}^{\mathcal{H}}\) is the bias in hyperbolic space \(\mathcal{H}_c^n\) with curvature \(c\) and the parallel transport \(P_{o \to \mathbf{x}^{\mathcal{H}}}^c\) is the function from the origin \(o\) to the hyperbolic point \(\mathbf{x}^{\mathcal{H}}\) under curvature \(c\). By using the method in Section \ref{mogan}, we calculate the attention-weighted representation:

\begin{equation}
\label{umohg}
    h_{i}^{l, \mathcal{H}, k} = \sigma \sum_{j=1}^{n} \left( \alpha_{ij}^{k} \left( W_{a, k}^{l, \mathcal{H}} x_{j}^{l, \mathcal{H}} + \epsilon_{a, k}^{l, \mathcal{H}} \right) \right),
\end{equation}
and apply the method in Section \ref{attention}:
\begin{equation}
    h_{i}^{l, \mathcal{H}} = \sum_{k=1}^{K} \mathrm{softmax}\left( s_{i}^{k, l, \mathcal{H}} \right) \cdot h_{i}^{k, l, \mathcal{H}}.
\end{equation}
These yield multi-order aggregated features in hyperbolic space, which are then mapped to a new hyperbolic space using the exponential function:

\begin{equation}
\label{umohgcaH}
    \mathcal{H} = \exp_{o}^{c} \left( h_{i}^{l, \mathcal{H}} \right),
\end{equation}
and its counterpart for negative sampling,

\begin{equation}
\label{umohgcaH'}
    \mathcal{H}' = \exp_{o}^{c} \left( h_{i}^{'l, \mathcal{H}} \right).
\end{equation}

\subsubsection{Contrastive Loss}
Before expressing in hyperbolic space, we map the features to the tangent space with Equation \ref{logP} and \ref{logL}, resulting in:

\begin{equation}
    \mathcal{X} = \log^{c}_{o} \left( \mathcal{H} \right),
\end{equation}
and
\begin{equation}
    \mathcal{X}' = \log^{c}_{o} \left( \mathcal{H}' \right).
\end{equation}
the contrastive loss function is defined as:
\begin{multline}
\label{lossU}
    \mathcal{L}_{MOUHGA} = \frac{1}{N + M} \Biggl( 
    \sum_{i=1}^{N} \mathcal{X} \bigl[ \log \mathcal{D} \bigl( e_{i}, \mathcal{Y} \bigr) \bigr] \\
    + \sum_{j=1}^{M} \mathcal{X}' \bigl[ \log \bigl( 1 - \mathcal{D} ( e'_{j}, \mathcal{Y} ) \bigr) \bigr]
\Biggr)
\end{multline}
, where \( \mathcal{Y} = \mathcal{R}(\mathcal{X}) \) is derived using a readout function \(\mathcal{R}(\cdot)\) defined as:

\begin{equation}
    \mathcal{R}(\mathcal{X}) = \sigma \left( \frac{1}{N} \sum_{i=1}^{N} e_i \right),
\end{equation}
where \(e_i \in \mathcal{X}\). Finally, a discriminator \(\mathcal{D}(\cdot)\) \cite{oord2018representation} maximizes the mutual information, formulated as:

\begin{equation}
    \mathcal{D}(e_i, \mathcal{Y}) = \rho \left( e_i \mathcal{W} \mathcal{Y} \right)
\end{equation}
, where there \(\mathcal{W}\) is a learnable scoring matrix, and \(\rho\) denotes the nonlinear logistic sigmoid function.

\subsection{Multi-Order Supervised Hyperbolic Graph Convolution and Aggregated Model for Social Event Detection}
\label{SMOHGCA}
\subsubsection{Overall Framework}

\begin{figure}
    \centering
    \fbox{\includegraphics[width=0.45\textwidth]{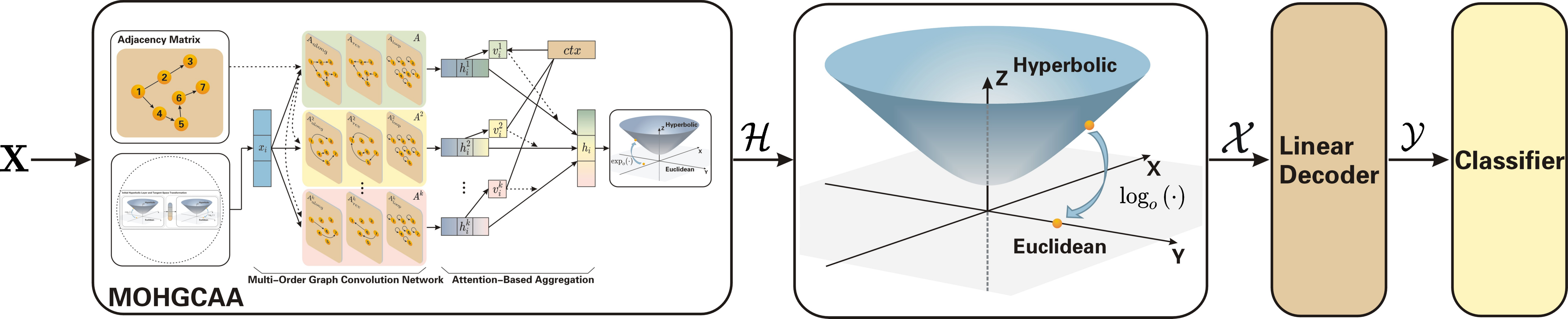}}
    \caption{The framework of the Multi-Order Hyperbolic Graph Convolution and Aggregated Model for Social Event Detection (MOHGCAASED) model. \(\mathbf{X} \) represents the node characteristics, \(\mathcal{H} \) signifies hyperbolic features subsequent to hyperbolic multi aggregation, \(\mathcal{X} \) indicates Euclidean space features following the logarithmic transformation, and \(\mathcal{Y} \) constitutes the ultimate feature representation.}
    \label{fig7}
\end{figure}

To validate the scalability and robustness of our proposed method, we applied the multi-order hyperbolic graph convolution aggregation to supervised learning tasks. The overall framework is illustrated in Figure \ref{fig7}. Specifically, the node features represented in Euclidean space are first initialized and mapped into hyperbolic space using the logarithmic map (Equation \ref{logP} and Equation \ref{logL}). Subsequently, the features are projected onto the tangent space at the origin o via the exponential map (Equation \ref{expP} and Equation \ref{expL}) to perform multi-order convolution aggregation. After aggregation, the updated features are mapped back into hyperbolic space using the logarithmic map (Equation \ref{logP} and Equation \ref{logL}). Finally, the processed features are passed through a linear decoder and output to the classifier for loss computation.
\subsubsection{Multi-Order Graph Convolution and Aggregation in Hyperbolic Space and Decoder Design}
We utilized the encoding approach proposed in Section \ref{MGCA} to obtain representations in hyperbolic space:
\begin{equation}
    \mathcal{H} = \exp_{o}^{c} \left( h_{i}^{l, \mathcal{H}} \right).
\end{equation}
We then project these embeddings back to Euclidean space using the logarithmic map (Equation \ref{logP} or Equation \ref{logL}):
\begin{equation}
    \mathcal{X} = \log^{c}_{o} \left( \mathcal{H} \right).
\end{equation}
After the projection, we compute the final linear decoding in Euclidean space as:

\begin{equation}
    \mathcal{Y} = \mathbf{W} \cdot \mathcal{X} + \mathbf{b}.
\end{equation}
Finally, we obtained the hyperbolic space feature representations of the nodes after multi-order convolution aggregation.
\subsubsection{Classification and Loss Function}
We define our task as a multi-node classification problem, where the given set  \(\mathcal{Y} = \{y_i \in \mathbb{R} | 1 \leq i \leq N\} \) represents \( N \)  message labels. The predicted probability for each class  \(p_i\)  is computed using the softmax function:

\begin{equation}
    p_i = \frac{e^{y_i}}{\sum_{j=1}^n e^{y_j}}.
\end{equation}
Finally, we minimize the cross-entropy loss function for training:
\begin{equation}
    \mathcal{L}_{MOHGCA} = -\sum_{i=1}^n l_i \log(p_i),
\end{equation}

where \(p_i\) is the predicted probability for class \(i\), and \(l_i\) represents the ground truth label.
\section{Experiments}
The experimental design was structured to address the following key research questions:
\begin{itemize}
    \item Q1: Does the proposed method outperform foundational approaches in prior works?
    \item Q2: How do variations in the number of node relation attention aggregation steps and dimensional settings influence experimental outcomes?
    \item Q3: Can experiments with multi-order configurations and dimensional settings empirically demonstrate that hyperbolic space outperforms Euclidean space?
    \item Q4: What are the effects of different models within the hyperbolic space on the performance of the proposed method?
\end{itemize}
These questions aim to provide a comprehensive evaluation of our model’s effectiveness, robustness, and theoretical advantages over existing methodologies.

\subsection{Datasets}
The datasets used in this experiment include:
Our model primarily targets social media data, so we selected the real-world Twitter \cite{mcminn2013building} dataset as our primary dataset. However, since the Twitter dataset is a large-scale dataset, contrastive learning with it may lead to out-of-memory errors. To address this, we used a balanced, smaller “mini-Twitter” dataset derived from the original dataset. Additionally, we used the Cora and the Citeseer \cite{sen2008collective} datasets to evaluate the effectiveness of modeling in hyperbolic space. Table \ref{tab:The statistics of datasets} provides an overview of these datasets.
\begin{table}[]
	\centering
    \caption{The statistics of datasets}
	\begin{tabular}{cccc}
		\hline
		Dataset              & \(\#\) of Classes      & \(\#\) of Nodes        & \(\#\) of Features     \\ \hline
		mini-Twitter         & 15                   & 3,000                & 302                  \\
		Twitter              & 503                  & 68,841               & 302                  \\
		Cora                 & 7                    & 2,708                & 1,433                \\
		Citeseer             & 6                    & 3,327                & 3,703                \\ \hline
		\multicolumn{1}{l}{} & \multicolumn{1}{l}{} & \multicolumn{1}{l}{} & \multicolumn{1}{l}{}
	\end{tabular}
	
	\label{tab:The statistics of datasets}
\end{table}
\subsection{MOHGCAA Experiments in Unsupervised Settings}
In this subsection, we evaluate the MOHGCAA model under unsupervised settings. Section \ref{baselineofU} introduces the baseline experiments, and Section \ref{psofU} provides a detailed description of the experimental environment and parameter settings. The remainder of this section addresses questions Q1 and Q2 with respect to the Multi-Order Supervised Hyperbolic Graph Convolution and Aggregated Model for Social Event Detection (MOUHGCAASED) model.

\subsubsection{Baseline Models}
\label{baselineofU}
To evaluate our model, we conducted extensive experiments using the mini-Twitter, Cora, and Citeseer datasets. We compared the performance against several state-of-the-art baselines, as detailed below:
\begin{itemize}
    \item DGI \cite{veli2017a}: A single-branch graph contrastive learning model. It generates negative samples by either augmenting data or corrupting the original graph structure, excelling in unsupervised learning of node representations through mutual information maximization.
    \item GraphCL \cite{you2020a}: A dual-branch graph contrastive learning framework. It performs augmentations on the input graph to generate two distinct views, capturing structural and semantic similarities to enhance representation learning.
    \item GCN \cite{kipf2016a}: A foundational graph neural network model leveraging spectral graph theory to generalize convolution operations to graph structures. By utilizing graph Laplacians, GCN performs layer-wise aggregation of features from local neighborhoods, enabling efficient semi-supervised learning.
    \item HGCN \cite{chami2019a}: An extension of GCN to hyperbolic space, designed for hierarchical and complex graph data. HGCN capitalizes on the unique properties of hyperbolic geometry to represent hierarchical structures with exponential capacity, achieving superior results in tasks like node classification and link prediction.
    \item HNN \cite{ganea2018a}: A hyperbolic neural network that operates directly in hyperbolic space to model hierarchical and structured datasets. By redefining core neural operations in hyperbolic geometry, HNN provides enhanced performance in classification and representation learning of complex data.
    \item HyboNet \cite{chen2021a}: A sophisticated hyperbolic neural framework fully utilizing the Lorentz model of hyperbolic geometry. HyboNet avoids inefficient mappings between Euclidean and hyperbolic spaces, incorporating hyperbolic rotations and attention mechanisms to achieve compact, powerful embeddings for relational and hierarchical data.
    \item UHSED \cite{qiu2024a}: An unsupervised hyperbolic graph-based model for social event detection, specifically tailored for heterogeneous and unlabelled social media datasets. By constructing a unified social message graph and employing contrastive learning, UHSED captures both semantic and structural information effectively.
\end{itemize}
The performance of these models was benchmarked on multiple tasks, highlighting the strengths and limitations of each in various contexts. This comparison underscores the advancements introduced by our proposed approach.

\subsubsection{Parameter Settings}
\label{psofU}
All the experiments with the model were conducted on an NVIDIA GeForce RTX 3090 GPU with AMD EPYC 7302 64‑core CPU processors. For other experimental hyperparameters, please refer to Table \ref{tab:pofUn}
\begin{table}[]
	\centering
    \caption{Parameters of MOHGCAA in Unsupervised Settings}
	\begin{tabular}{ccll}
		\cmidrule{1-2}
		Parameter           & Value              &  &  \\ \cmidrule{1-2}
		Mulit-order         & 4                  &  &  \\
		Hidden layer        & 1                  &  &  \\
		Hidden dimension    & 512                &  &  \\
		Drop rate           & 10\%                &  &  \\
		Learning rate       & 0.1                &  &  \\
		Optimiser           & Adam               &  &  \\
		Activation function & ReLU               &  &  \\
		Augmentation method & Feature corruption &  &  \\ \cmidrule{1-2}
	\end{tabular}
	\label{tab:pofUn}
\end{table}
\subsubsection{Evaluation Metrics}
Since our proposed method is an unsupervised learning model, we adopted Micro-F1 and Macro-F1 scores to evaluate the accuracy of the detection results, following the methodology of prior studies \cite{veli2017a,you2020a}. These metrics were chosen for their robustness in assessing performance across imbalanced datasets, with Micro-F1 capturing overall classification performance and Macro-F1 emphasizing class-specific performance.

\subsubsection{Performance Comparison of MOHGCAA in Unsupervised Scenarios (Answer Q1)}
The performance of our method in unsupervised settings, evaluated on the mini-Twitter, Cora, and Citeseer datasets, is summarized in Table \ref{tab:performance_comparison}. Overall, our approach consistently outperforms existing models based on Euclidean and hyperbolic spaces. The results across datasets from different domains demonstrate the significant effectiveness of our method.

\begin{table*}
    \centering
    \caption{Performance Comparison of Different Models on Various Datasets}
    \begin{tabular}{ccccccc}
        \toprule
        \multirow{2}{*}{\textbf{Models}} & \multicolumn{2}{c}{\textbf{mini-twitter}} & \multicolumn{2}{c}{\textbf{cora}} & \multicolumn{2}{c}{\textbf{citeseer}} \\ 
        \cmidrule(lr){2-3} \cmidrule(lr){4-5} \cmidrule(lr){6-7}
        & Micro-F1 & Macro-F1 & Micro-F1 & Macro-F1 & Micro-F1 & Macro-F1 \\ 
        \midrule
        DGI       & 0.1714  & 0.1413  & 0.8077  & 0.7922  & 0.6885  & 0.6518  \\ 
        GraphCL   & 0.1517  & 0.1302  & 0.6993  & 0.6837  & 0.6377  & \textbf{0.7971}  \\ 
        GCN       & 0.4774  & 0.4850  & 0.7960  & 0.7883  & 0.5435  & 0.4496  \\ 
        HNN       & 0.3896  & 0.3444  & 0.3610  & 0.2544  & 0.5202  & 0.4815  \\ 
        HGCN      & \underline{0.5451}  & \underline{0.5526}  & 0.7920  & 0.7878  & 0.7038  & 0.6600  \\ 
        HyboNet   & 0.4111  & 0.4079  & 0.8070  & 0.7991  & 0.5562  & 0.5283  \\ 
        UHSED     & 0.5288  & 0.5266  & \underline{0.8314}  & \underline{0.8203}  & \underline{0.7081}  & 0.6458  \\ 
        MOHGAA & \textbf{0.6062}  & \textbf{0.5980}  & \textbf{0.8543}  & \textbf{0.8425}  & \textbf{0.7315}  & 0.6707  \\ 
        \bottomrule
    \end{tabular}
    \label{tab:performance_comparison}
\end{table*}

\subsubsection{Impact of Different Multi-Order Steps on Model Performance (Answer Q2)}
\begin{figure*}
    \centering
    \fbox{\includegraphics[width=0.9\textwidth]{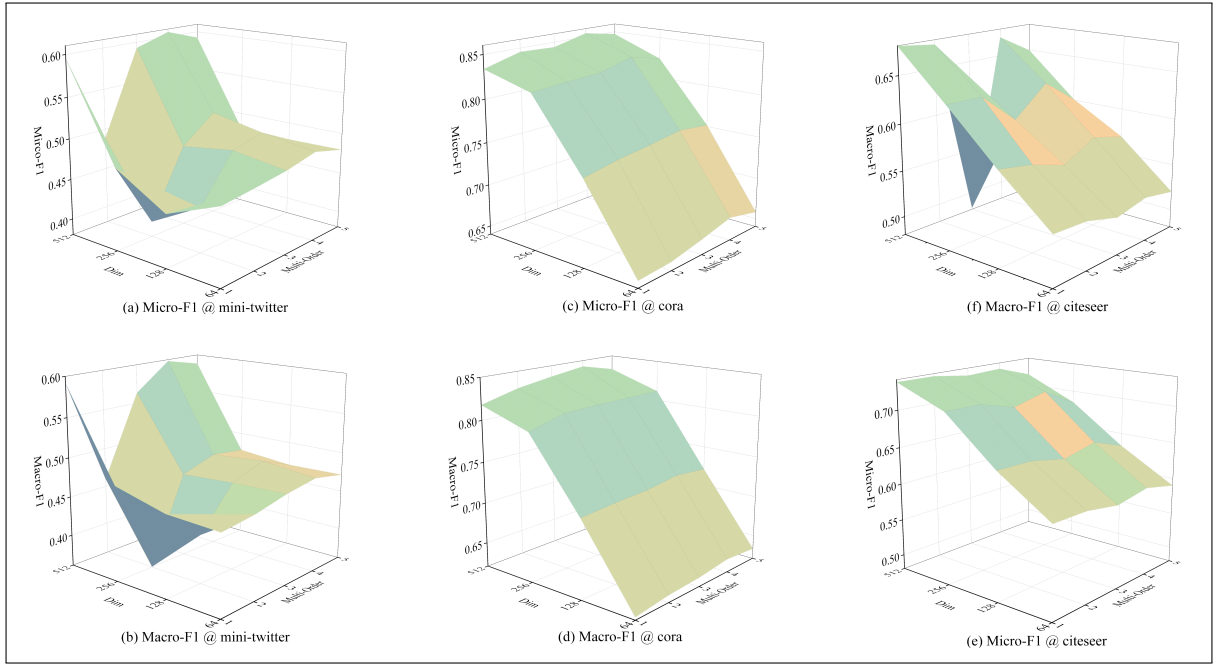}}
    \caption{Different Multi-Orders and Dimension of MOHGCAA in Unsupervised Settings}
    \label{fig8}
\end{figure*}

We conducted ablation experiments to evaluate the impact of different orders and node feature dimensions in hyperbolic space on our method’s performance. The results reveal that higher-dimensional node features generally yield better performance compared to lower-dimensional ones. However, the performance does not increase proportionally with the number of multi-order steps, as per the core idea of multi-order relational attention aggregation. This is because, after a certain number of steps, the relationships become sparser, leading to a diminished rate of information aggregation until the optimal performance is achieved for a specific dataset. As illustrated in Figure \ref{fig8}, both in high-dimensional and low-dimensional settings, the best performance is consistently achieved at the fourth aggregation step.

\subsection{MOHGCAA Experiments in Supervised Settings}
This section presents the experimental results of MOHGCAA. Section \ref{baselineofSUP} introduces the baseline experiments, while Section \ref{paraofS} describes the experimental setup and parameter configurations. The remaining subsections address the key research questions Q1 and Q2.

\subsubsection{Baseline Models}
\label{baselineofSUP}
To validate the robustness of our method, we conducted experiments in supervised scenarios. Intuitive experiments were performed to compare results on the Twitter dataset using various models, including both traditional classic models and state-of-the-art (SOTA) models currently under research. The comparisons include models in Euclidean space and hyperbolic space-based models:
\begin{itemize}
    \item Word2vec \cite{mikolov2013a}:An approach employing universal embedding representation for the social event identification.
    \item LDA \cite{blei2003a}: A traditional statistical approach is employed for the social event detection.
    \item WMD \cite{kusner2015a}: A similarity-based measurement method for detecting social events by computing the similarity between messages.
    \item BERT \cite{devlin2018a}: A robust language representation model that plays a crucial role in numerous state-of-the-art social event detection models.
    \item KPGNN \cite{cao2021a}: A social event detection model based on heterogeneous information networks (HINs), primarily designed for incremental social event detection. Its offline performance slightly outperforms that of PP-GCN.
    \item FinEvent \cite{peng2022a}: A social event detection model that leverages incremental and cross-lingual social messages, employing reinforcement learning to enhance detection performance.
    \item HSED \cite{qiu2024a}: A hyperbolic space-based social event detection model.
    \item HNN \cite{ganea2018a}: Another hyperbolic neural network model applied for social event detection.
\end{itemize}
These experiments aim to comprehensively evaluate the performance of MOHGCAA in supervised scenarios and its robustness across different types of datasets and modeling approaches.

\subsubsection{Parameter Settings}
\label{paraofS}
All the experiments with the model were conducted on an NVIDIA GeForce RTX 3090 GPU with AMD EPYC 7302 64‑core CPU processors. For other experimental hyperparameters, please refer to Table \ref{tab:pofSup}.
\begin{table}[]
	\centering
    \caption{Parameters of MOHGCAA in Supervised Settings}
	\begin{tabular}{ccll}
		\cline{1-2}
		Parameter           & Value &  &  \\ \cline{1-2}
		Multi-order         & 2     &  &  \\
		Hidden layer        & 2     &  &  \\
		Hidden dimension    & 512   &  &  \\
		Traing rate         & 70\%   &  &  \\
		Test rate           & 20\%   &  &  \\
		Validataion rate    & 10\%   &  &  \\
		Learning rate       & 0.1   &  &  \\
		Optimiser           & Adam  &  &  \\
		Activation function & ReLU  &  &  \\ \cline{1-2}
	\end{tabular}
	\label{tab:pofSup}
\end{table}
\subsubsection{Evaluation Metrics}
To ensure the robustness of our experiments, even in specialized domains, we adopted standard evaluation metrics commonly used in event detection within the news domain. These metrics, namely NMI [52], AMI [53], and ARI [53], have been widely employed to evaluate social event detection models [2], [17]. These metrics provide a fair basis for comparison and ensure the reliability of our experimental results.

\subsubsection{Performance Comparison of the MOHGCAA Model (Answer Q1)}
The experimental results comparing our method with other models in non-Euclidean and hyperbolic spaces on the Twitter dataset are summarized in Table \ref{tab:Performance Comparison of Different Models on Twitter data}. Overall, our approach consistently outperforms the baseline models. In terms of accuracy, the predicted results closely align with the ground truth labels.
We also compared MOHGCAA with HSED and HNN. While the best results achieved by these models appear similar, our ablation studies revealed performance differences across varying multi-hop steps and dimensions. These distinctions, and their implications, will be discussed in detail in Section \ref{iDMOS}.
\begin{table}[]
	\centering
    \caption{Performance Comparison of Different Models on Twitter dataset}
	\begin{tabular}{cccccc}
		\toprule
		\multicolumn{2}{c}{Models}                                    & ACC                       & NMI   & AMI   & ARI   \\ \midrule
		\multirow{6}{*}{euclidean}  & \multicolumn{1}{|c}{Word2vec} & \multicolumn{1}{|c}{0.32} & 0.41  & 0.12  & 0.02  \\
		                            & \multicolumn{1}{|c}{LDA}      & \multicolumn{1}{|c}{0.2} & 0.28  & 0.04  & 0.01  \\
		                            & \multicolumn{1}{|c}{WMD}      & \multicolumn{1}{|c}{-}    & 0.63* & 0.49*  & 0.06* \\
		                            & \multicolumn{1}{|c}{BERT}     & \multicolumn{1}{|c}{0.51} & 0.61  & 0.41  & 0.07  \\
		                            & \multicolumn{1}{|c}{KPGNN}    & \multicolumn{1}{|c}{-}    & 0.69*  & 0.5* & 0.21* \\
		                            & \multicolumn{1}{|c}{FinEvent} & \multicolumn{1}{|c}{-}    & 0.79*  & 0.69* & 0.48* \\ \midrule
		\multirow{3}{*}{hyperbolic} & \multicolumn{1}{|c}{HNN}      & \multicolumn{1}{|c}{\underline{0.89}} & 0.91  & 0.71  & 0.87  \\
		                            & \multicolumn{1}{|c}{HSED}     & \multicolumn{1}{|c}{\underline{0.89}} & \underline{0.92}  & \underline{0.73}  & \underline{0.88}  \\
		                            & \multicolumn{1}{|c}{MOHGCAA}   & \multicolumn{1}{|c}{\textbf{0.91}} & \textbf{0.93}  & \textbf{0.75}  & \textbf{0.9}   \\ \bottomrule
	\end{tabular}
	
	\label{tab:Performance Comparison of Different Models on Twitter data}
\end{table}

\subsubsection{Impact of Dimensionality and Multi-Order Steps on Model Performance (Answer Q2)}
\label{iDMOS}
\begin{figure*}
    \centering
    \fbox{\includegraphics[width=\textwidth]{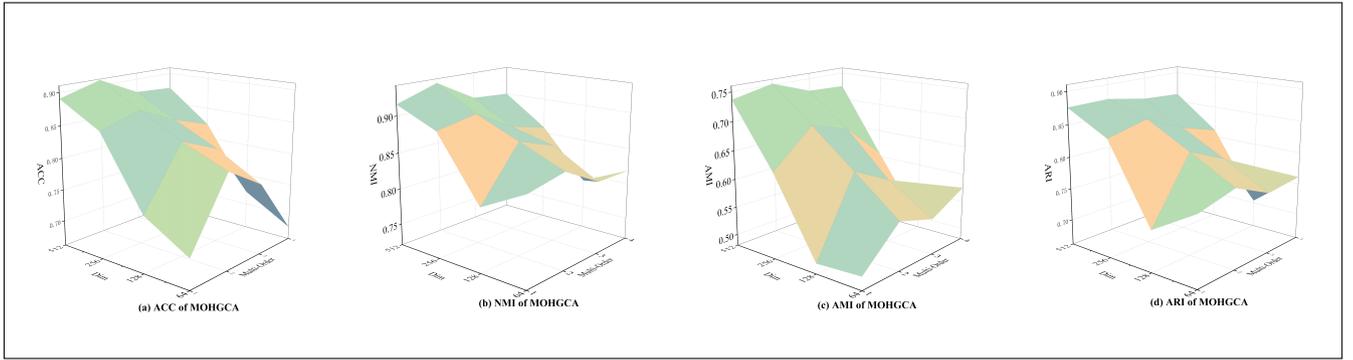}}
    \caption{Different Multi-Orders and Dimension of MOHGCAA in Supervised Settings}
    \label{fig9}
\end{figure*}

Following the same ablation study methodology used for the unsupervised models in the previous section, we conducted comparative experiments on different multi-order attention aggregation steps and node feature dimensions in hyperbolic space.
Our findings reveal that, higher-dimensional node features encode more comprehensive node information. Regarding the number of multi-order steps, consistent with the findings from the previous ablation study, the best overall performance is achieved at the second order. Readers can refer to Figure \ref{fig9} for detailed results and visualizations.
\subsection{Discussion}
\subsubsection{Performance Comparison in Euclidean and Hyperbolic Spaces (Answer Q3)}
\begin{figure*}
    \centering
    \fbox{\includegraphics[width=0.9\textwidth]{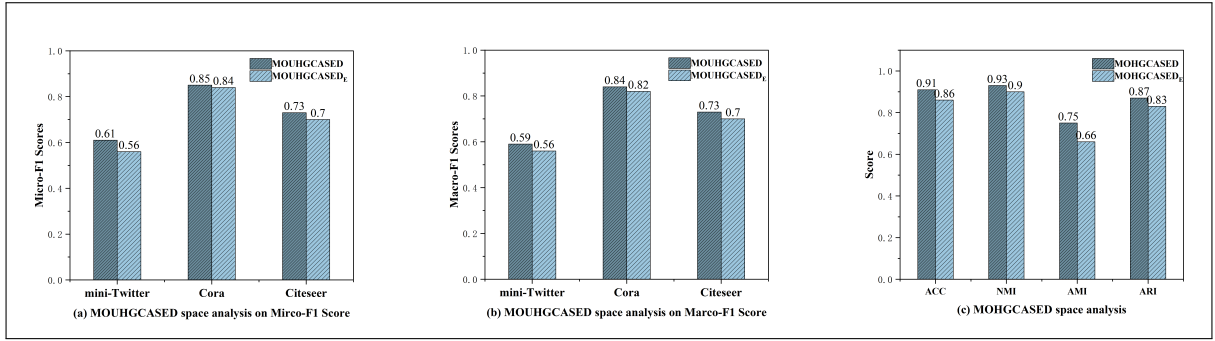}}
    \caption{Comparison of Euclidean and Hyperbolic Spaces in Unsupervised and Supervised Settings. (a) Micro-F1 scores for Euclidean and hyperbolic spaces in the unsupervised scenario. (b) Macro-F1 scores for Euclidean and hyperbolic spaces in the unsupervised scenario. (c) Various metric scores for Euclidean and hyperbolic spaces in the supervised scenario.}
    \label{fig10}
\end{figure*}

To evaluate the effectiveness of our method across two distinct spaces, we compared the performance in Euclidean and hyperbolic spaces under both unsupervised settings. The results are presented in Figures \ref{fig10}. In these comparisons: \(MOUHGASED\) and \(MOHGASED\) represents multi-order attention aggregation in hyperbolic space. \(MOUHGASED_E\) and \(MOHGASED_E\) denotes attention aggregation in Euclidean space. Our findings demonstrate that hyperbolic space consistently outperforms Euclidean space across both general event detection datasets and specialized datasets, regardless of the model type.

\subsubsection{Performance Comparison Among Hyperbolic Space Models (Answer Q4)}
\begin{figure*}
    \centering
    \fbox{\includegraphics[width=0.9\textwidth]{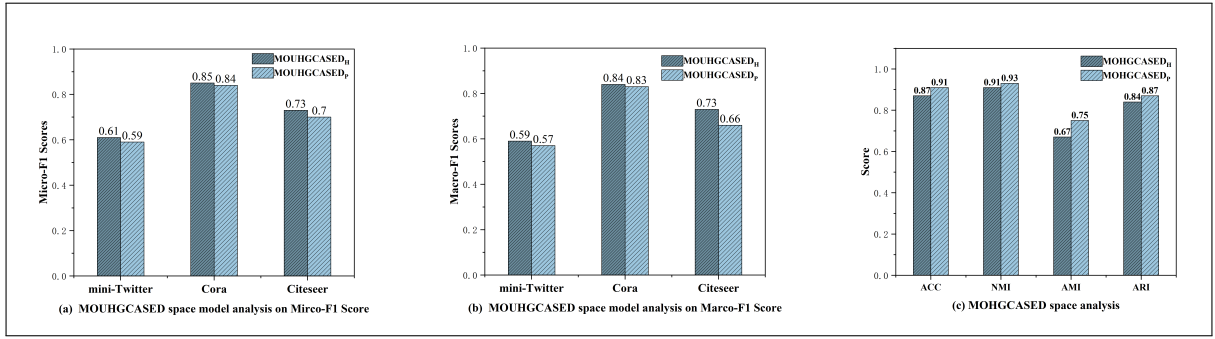}}
    \caption{Comparison in different Hyperbolic Spaces in Unsupervised and Supervised Settings. (a) Micro-F1 scores in the unsupervised scenario. (b) Macro-F1 scores in the unsupervised scenario. (c) Various metric scores in the supervised scenario.}
    \label{fig11}
\end{figure*}

As introduced in Section \ref{Perliminary}, two popular hyperbolic space models are the Poincar´e ball model and the Lorentz model. Although these models are mathematically equivalent, our experiments reveal subtle performance differences depending on the dataset and whether the method is unsupervised or supervised. As shown in Figures \ref{fig11}, the Lorentz model exhibits slight advantages in unsupervised settings, while the Poincar´e ball model performs marginally better in supervised scenarios. These results highlight nuanced distinctions in their applicability to different tasks and datasets.
\section{Conclusion}
In this paper, we addressed the challenges of social event detection (SED) by proposing a novel framework, Multi-Order Hyperbolic Graph Convolution with Aggregated Attention (MOHGCAA). Our approach effectively captures the hierarchical and higher-order relationships in event detection data, overcoming the limitations of existing methods in Euclidean and hyperbolic spaces. Experimental results on multiple datasets demonstrated that MOHGCAA consistently achieves superior performance under both supervised and unsupervised settings, validating its effectiveness and robustness. These findings highlight the framework’s capability to enhance SED applications across diverse scenarios, offering valuable insights for businesses and public services alike. Moving forward, our work opens new possibilities for leveraging hyperbolic graph convolution networks in other hierarchical data domains. Future research may focus on further optimizing the framework for scalability and exploring its potential in real-time event detection and other dynamic graph-based tasks.
\bibliographystyle{unsrt}
\bibliography{references}

\begin{thebibliography}{10}

\bibitem{fedoryszak2019a}
M.~Fedoryszak, B.~Frederick, V.~Rajaram, and C.~Zhong.
\newblock Real-time event detection on social data streams.
\newblock In {\em Proceedings of the 25th ACM SIGKDD international conference
  on knowledge discovery \& data mining}, page 2774–2782, 2019.

\bibitem{peng2019a}
H.~Peng, J.~Li, Q.~Gong, Y.~Song, Y.~Ning, K.~Lai, and P.S. Yu.
\newblock Fine-grained event categorization with heterogeneous graph
  convolutional networks, 2019.
\newblock arXiv preprint arXiv:1906.0458.

\bibitem{zhou2020a}
H.~Zhou, H.~Yin, H.~Zheng, and Y.~Li.
\newblock A survey on multi-modal social event detection.
\newblock {\em Knowledge-Based Systems}, 195:105695, 2020.

\bibitem{cao2021a}
Y.~Cao, H.~Peng, J.~Wu, Y.~Dou, J.~Li, and P.S. Yu.
\newblock Knowledge-preserving incremental social event detection via
  heterogeneous gnns.
\newblock In {\em Proceedings of the Web Conference 2021}, page 3383–3395,
  2021.

\bibitem{cui2021a}
W.~Cui, J.~Du, D.~Wang, F.~Kou, and Z.~Xue.
\newblock Mvgan: Multi-view graph attention network for social event detection.
\newblock {\em ACM Transactions on Intelligent Systems and Technology (TIST},
  12(3):1–24, 2021.

\bibitem{afyouni2022a}
I.~Afyouni, Z.~Al~Aghbari, and R.A. Razack.
\newblock Multi-feature, multi-modal, and multi-source social event detection:
  A comprehensive survey.
\newblock {\em Information Fusion}, 79:279–308, 2022.

\bibitem{qian2023a}
S.~Qian, H.~Chen, D.~Xue, Q.~Fang, and C.~Xu.
\newblock Open-world social event classification.
\newblock In {\em Proceedings of the ACM Web Conference 2023}, page
  1562–1571, 2023.

\bibitem{qiu2024a}
Z.~Qiu, J.~Wu, J.~Yang, X.~Su, and C.C. Aggarwal.
\newblock Heterogeneous social event detection via hyperbolic graph
  representations.
\newblock {\em IEEE Transactions on Big Data}, 2024.

\bibitem{cao2024a}
Y.~Cao, H.~Peng, Z.~Yu, and S.Y. Philip.
\newblock Hierarchical and incremental structural entropy minimization for
  unsupervised social event detection.
\newblock {\em Proceedings of the AAAI Conference on Artificial Intelligence},
  38(8):8255–8264, 2024.

\bibitem{guo2024a}
Y.~Guo, Z.~Zang, H.~Gao, X.~Xu, R.~Wang, L.~Liu, and J.~Li.
\newblock Unsupervised social event detection via hybrid graph contrastive
  learning and reinforced incremental clustering.
\newblock {\em Knowledge-Based Systems}, 284:111225, 2024.

\bibitem{mikolov2013a}
T.~Mikolov, K.~Chen, G.~Corrado, and J.~Dean.
\newblock Efficient estimation of word representations in vector space.
\newblock arXiv preprint arXiv:1301.3781.

\bibitem{ahn2006a}
D.~Ahn.
\newblock The stages of event extraction.
\newblock In {\em Proceedings of the Workshop on Annotating and Reasoning about
  Time and Events}, page 1–8, 2006.

\bibitem{liao2011a}
S.~Liao and R.~Grishman.
\newblock Acquiring topic features to improve event extraction: in pre-selected
  and balanced collections.
\newblock In {\em Proceedings of the International Conference Recent Advances
  in Natural Language Processing 2011}, page 9–16, 2011.

\bibitem{patwardhan2009a}
S.~Patwardhan and E.~Riloff.
\newblock A unified model of phrasal and sentential evidence for information
  extraction.
\newblock In {\em Proceedings of the 2009 conference on empirical methods in
  natural language processing}, page 151–160, 2009.

\bibitem{nguyen2015a}
T.H. Nguyen and R.~Grishman.
\newblock Event detection and domain adaptation with convolutional neural
  networks.
\newblock In {\em Proceedings of the 53rd Annual Meeting of the Association for
  Computational Linguistics and the 7th International Joint Conference on
  Natural Language Processing (Volume 2: Short Papers}, page 365–371, 2015.

\bibitem{chen2015a}
Y.~Chen, L.~Xu, K.~Liu, D.~Zeng, and J.~Zhao.
\newblock Event extraction via dynamic multi-pooling convolutional neural
  networks.
\newblock In {\em Proceedings of the 53rd Annual Meeting of the Association for
  Computational Linguistics and the 7th International Joint Conference on
  Natural Language Processing (Volume 1: Long Papers}, page 167–176, 2015.

\bibitem{zhang2016a}
Z.~Zhang, W.~Xu, and Q.~Chen.
\newblock Joint event extraction based on skip-window convolutional neural
  networks.
\newblock In {\em Natural Language Understanding and Intelligent Applications:
  5th CCF Conference on Natural Language Processing and Chinese Computing,
  NLPCC 2016, and 24th International Conference on Computer Processing of
  Oriental Languages, ICCPOL 2016}, volume Proceedings 24, page 324–334,
  Kunming, China, 2016. Springer International Publishing.

\bibitem{li2018a}
L.~Li, Y.~Liu, and M.~Qin.
\newblock Extracting biomedical events with parallel multi-pooling
  convolutional neural networks.
\newblock {\em IEEE/ACM transactions on computational biology and
  bioinformatics}, 17(2):599–607, 2018.

\bibitem{kodelja2019a}
D.~Kodelja, R.~Besançon, and O.~Ferret.
\newblock Exploiting a more global context for event detection through
  bootstrapping.
\newblock In {\em European conference on information retrieval}, page
  763–770, Cham, 2019. Springer International Publishing.

\bibitem{hochreiter1997a}
S.~Hochreiter.
\newblock Long short-term memory.
\newblock In {\em Neural Computation MIT-Press}. 1997.

\bibitem{ghaeini2018a}
R.~Ghaeini, X.Z. Fern, L.~Huang, and P.~Tadepalli.
\newblock Event nugget detection with forward-backward recurrent neural
  networks.
\newblock arXiv preprint arXiv:1802.05672.

\bibitem{chen2016a}
Y.~Chen, S.~Liu, S.~He, K.~Liu, and J.~Zhao.
\newblock Event extraction via bidirectional long short-term memory tensor
  neural networks.
\newblock In {\em Chinese Computational Linguistics and Natural Language
  Processing Based on Naturally Annotated Big Data: 15th China National
  Conference, CCL 2016, and 4th International Symposium, NLP-NABD 2016}, volume
  Proceedings 4, page 190–203, Yantai, China, 2016. Springer International
  Publishing.

\bibitem{sha2018a}
L.~Sha, F.~Qian, B.~Chang, and Z.~Sui.
\newblock Jointly extracting event triggers and arguments by dependency-bridge
  rnn and tensor-based argument interaction.
\newblock {\em Proceedings of the AAAI conference on artificial intelligence},
  32(1)).

\bibitem{li2019a}
D.~Li, L.~Huang, H.~Ji, and J.~Han.
\newblock Biomedical event extraction based on knowledge-driven tree-lstm.
\newblock In {\em Proceedings of the 2019 Conference of the North American
  Chapter of the Association for Computational Linguistics: Human Language
  Technologies}, volume~1, page 1421–1430.

\bibitem{kipf2016a}
T.N. Kipf and M.~Welling.
\newblock Semi-supervised classification with graph convolutional networks.
\newblock arXiv preprint arXiv:1609.0290.

\bibitem{nguyen2018a}
T.~Nguyen and R.~Grishman.
\newblock Graph convolutional networks with argument-aware pooling for event
  detection.
\newblock {\em Proceedings of the AAAI Conference on Artificial Intelligence},
  32(1)).

\bibitem{liu2018a}
X.~Liu, Z.~Luo, and H.~Huang.
\newblock Jointly multiple events extraction via attention-based graph
  information aggregation.
\newblock arxiv preprint arxiv:1809.09078.

\bibitem{yan2019a}
Haoran Yan, Xiaolong Jin, Xiangbin Meng, Jiafeng Guo, and Xueqi Cheng.
\newblock Event detection with multi-order graph convolution and aggregated
  attention.
\newblock In {\em Proceedings of the 2019 conference on empirical methods in
  natural language processing and the 9th international joint conference on
  natural language processing (emnlp-ijcnlp}, page 5766–5770.

\bibitem{lai2020a}
V.D. Lai, T.N. Nguyen, and T.H. Nguyen.
\newblock Event detection: Gate diversity and syntactic importance scoresfor
  graph convolution neural networks.
\newblock arxiv preprint arxiv:2010.14123.

\bibitem{cui2020a}
S.~Cui, B.~Yu, T.~Liu, Z.~Zhang, X.~Wang, and J.~Shi.
\newblock Edge-enhanced graph convolution networks for event detection with
  syntactic relation.
\newblock arxiv preprint arxiv:2002.10757.

\bibitem{dutta2021a}
S.~Dutta, L.~Ma, T.K. Saha, D.~Lu, J.~Tetreault, and A.~Jaimes.
\newblock Gtn-ed: Event detection using graph transformer networks.
\newblock arxiv preprint arxiv:2104.15104.

\bibitem{blei2003a}
D.M. Blei, A.Y. Ng, and M.I. Jordan.
\newblock Latent dirichlet allocation.
\newblock {\em Journal of machine Learning research}, 3(Jan):993–1022.

\bibitem{li20161a}
C.~Li, H.~Wang, Z.~Zhang, A.~Sun, and Z.~Ma.
\newblock Topic modeling for short texts with auxiliary word embeddings.
\newblock In {\em Proceedings of the 39th International ACM SIGIR conference on
  Research and Development in Information Retrieval}, page 165–174.

\bibitem{shi2018a}
T.~Shi, K.~Kang, J.~Choo, and C.K. Reddy.
\newblock Short-text topic modeling via non-negative matrix factorization
  enriched with local word-context correlations.
\newblock In {\em Proceedings of the 2018 world wide web conference}, page
  1105–1114.

\bibitem{liu2019a}
Q.~Liu, M.~Nickel, and D.~Kiela.
\newblock Hyperbolic graph neural networks.
\newblock In {\em Advances in neural information processing systems}, page~32.

\bibitem{peng2020a}
W.~Peng, J.~Shi, Z.~Xia, and G.~Zhao.
\newblock Mix dimension in poincaré geometry for 3d skeleton-based action
  recognition.
\newblock In {\em Proceedings of the 28th ACM International Conference on
  Multimedia}, page 1432–1440.

\bibitem{elsken2019a}
T.~Elsken, J.H. Metzen, and F.~Hutter.
\newblock Neural architecture search: A survey.
\newblock {\em Journal of Machine Learning Research}, 20(55):1–21.

\bibitem{bachmann2020a}
G.~Bachmann, G.~Bécigneul, and O.~Ganea.
\newblock Constant curvature graph convolutional networks.
\newblock In {\em International conference on machine learning}, page
  486–496. PMLR.

\bibitem{qiu2024b}
Z.~Qiu, C.~Ma, J.~Wu, and J.~Yang.
\newblock An efficient automatic meta-path selection for social event detection
  via hyperbolic space.
\newblock In {\em Proceedings of the ACM on Web Conference 2024}, page
  2519–2529.

\bibitem{oord2018representation}
Aaron van~den Oord, Yazhe Li, and Oriol Vinyals.
\newblock Representation learning with contrastive predictive coding.
\newblock {\em arXiv preprint arXiv:1807.03748}, 2018.

\bibitem{mcminn2013building}
Andrew~J McMinn, Yashar Moshfeghi, and Joemon~M Jose.
\newblock Building a large-scale corpus for evaluating event detection on
  twitter.
\newblock In {\em Proceedings of the 22nd ACM international conference on
  Information \& Knowledge Management}, pages 409--418, 2013.

\bibitem{sen2008collective}
Prithviraj Sen, Galileo Namata, Mustafa Bilgic, Lise Getoor, Brian Galligher,
  and Tina Eliassi-Rad.
\newblock Collective classification in network data.
\newblock {\em AI magazine}, 29(3):93--93, 2008.

\bibitem{veli2017a}
P.~Veličković, G.~Cucurull, A.~Casanova, A.~Romero, P.~Lio, and Y.~Bengio.
\newblock Graph attention networks. arXiv preprint arXiv:1710.10903.

\bibitem{you2020a}
Y.~You, T.~Chen, Y.~Sui, T.~Chen, Z.~Wang, and Y.~Shen.
\newblock Graph contrastive learning with augmentations.
\newblock {\em Advances in neural information processing systems},
  33:5812–5823.

\bibitem{chami2019a}
I.~Chami, Z.~Ying, C.~Ré, and J.~Leskovec.
\newblock Hyperbolic graph convolutional neural networks.
\newblock In {\em Advances in neural information processing systems}, page~32.

\bibitem{ganea2018a}
O.~Ganea, G.~Bécigneul, and T.~Hofmann.
\newblock Hyperbolic neural networks.
\newblock In {\em Advances in neural information processing systems}, page~31.

\bibitem{chen2021a}
W.~Chen, X.~Han, Y.~Lin, H.~Zhao, Z.~Liu, P.~Li, and J.~Zhou.
\newblock Fully hyperbolic neural networks.
\newblock arXiv preprint arXiv:2105.14686.

\bibitem{kusner2015a}
M.~Kusner, Y.~Sun, N.~Kolkin, and K.~Weinberger.
\newblock From word embeddings to document distances.
\newblock In {\em International conference on machine learning}, page
  957–966. PMLR.

\bibitem{devlin2018a}
J.~Devlin.
\newblock Bert: Pre-training of deep bidirectional transformers for language
  understanding.
\newblock arXiv preprint arXiv:1810.04805.

\bibitem{peng2022a}
H.~Peng, R.~Zhang, S.~Li, Y.~Cao, S.~Pan, and S.Y. Philip.
\newblock Reinforced, incremental and cross-lingual event detection from social
  messages.
\newblock {\em IEEE Transactions on Pattern Analysis and Machine Intelligence},
  45(1):980–998.

\end{thebibliography}

\appendix
\section{My Appendix}

\printcredits

\bio{figs/author1}
{Yao Liu} received the B.S. and M.S. degrees from School of Computer Science, Southwest Petroleum University, China, in 2007. He is currently pursuing the Ph.D. degree with the School of Computer Science, Universiti Sains Malaysia, Penang, Malaysia. He is currently an Associate Professor with the Department of Management and Media, The Engineering and Technology College of Chengdu University of Technology. His research interests include natural language processing and legal artificial intelligence.
\endbio

\bio{figs/author2}
{Tien-Ping Tan} received the Ph.D. degree from Université Joseph Fourier, France, in 2008. He is currently an Associate Professor with the School of Computer Sciences, Universiti Sains Malaysia. His research interests include automatic speech recognition, machine translation, and natural language processing.
\endbio

\bio{figs/author3}
{Zhilan Liu} received the B.S. degree from Chengdu University of Technology, China, in 2003, and the M.S. degree from Jiangnan University, China, in 2012. She is currently an Associate Professor with the Department of Art Design, The Engineering and Technology College of Chengdu University of Technology. Her research interests include environmental space design within the context of regional culture.
\endbio

\bio{figs/author4}
{Yuxin Li} received the master degree from University of Surrey, UK. Now, she works in School of Department of Management and Media, The Engineering and Technical College of Chengdu University of Technology. Her research interests include micro video production, audio-visual language, Journalism and communication
\endbio
\end{document}